\documentclass[12pt]{iopart}

\usepackage{iopams}  
\usepackage{harvard}
\usepackage{epsfig}
\usepackage{graphicx}

\begin{document}

\topical{Singular responses of spin-incoherent Luttinger liquids}

\author{Gregory A. Fiete}

\address{Department of Physics, The University of Texas at Austin, Austin, TX 78712  USA}
\ead{fiete@physics.utexas.edu}

\begin{abstract}
When a local potential changes abruptly in time, an electron gas responds by shifting to a new state which at long times is orthogonal to the one in the absence of the local potential.  This is known as Anderson's orthogonality catastrophe and it is relevant for the so-called X-ray edge or Fermi edge singularity, and for tunneling into an interacting one dimensional system of fermions.  It often happens that the finite frequency response of the photon absorption or the tunneling density of states exhibits a singular behavior as a function of frequency: $\left(\frac{\omega_{\rm th}}{\omega-\omega_{\rm th}}\right)^\alpha\Theta(\omega-\omega_{\rm th})$ where $\omega_{\rm th}$ is a threshold frequency and $\alpha$ is an exponent characterizing the singular response.  In this paper singular responses of spin-incoherent Luttinger liquids are reviewed.  Such responses most often do not fall into the familiar form above, but instead typically exhibit logarithmic corrections and display a much higher universality in terms of the microscopic interactions in the theory.  Specific predictions are made, the current experimental situation is summarized, and key outstanding theoretical issues related to spin-incoherent Luttinger liquids are highlighted.  

\end{abstract}

\maketitle

\section{Introduction}

Until a few years ago there was a common feeling in the condensed matter physics community that we had a very good understanding of the universal properties of one-dimensional systems of fermions through our knowledge of Luttinger liquid theory and its instabilities \cite{Giamarchi,Gogolin}.  One of the key results of Luttinger liquid theory \cite{haldane81} is that interacting electrons in one-dimension have separate dynamics in the spin and charge sectors at low energies.  These low energy collective spin and charge modes are bosonic and generally propagate with different velocities, with the spin mode typically being the slower of the two.  Luttinger liquid theory provides a particularly convenient theoretical framework for discussing the response of the electron gas to a time-dependent local potential, even in higher dimensions: The mathematical trick of ``bosonization" makes the evaluation of correlation functions relatively simple, and a potential local on the scale of the Fermi wavelength results in scattering of primarily $s$-wave type which effectively converts the problem to just one dimension (the radial coordinate).

A few years back it was realized that at finite temperatures a ``Luttinger liquid system" can take on qualitatively new behaviors in transport \cite{Matveev:prl04} and the single-particle Green's function \cite{cheianov03}.  Some of the results for the single-particle Green's function were anticipated in an earlier work \cite{Berkovich:jpa91}, but the many novel features that can arise due to finite temperatures were not fully appreciated until the more recent works mentioned above which really started proper the study of the spin-incoherent Luttinger liquid \cite{Fiete:rmp07}.  

The spin-incoherent regime of a one-dimensional interacting electron gas is defined in the following way.  Let us assume that at zero temperature our system is spin-charge separated with spin velocity $v_s$ and charge velocity $v_c$ such that $v_s \ll v_c$. If the interactions are strong (regardless of the range of interactions) it can be shown that $v_s$ can be {\em exponentially small} compared to $v_c$ \cite{Hausler:zpb96,Matveev:prb04,Fogler_exch:prb05}.  Defining a characteristic spin energy $E_s =\hbar v_s/a$ and a characteristic charge energy $E_c=\hbar v_c/a$ for mean particle spacing $a$, it is evidently clear that the spin energy can be exponentially suppressed relative to the charge energy: $E_s \ll E_c$.  An exponentially separated spin and charge energy makes it possible for the thermal energy to lie in the range $E_s \ll k_B T \ll E_c$ where $k_B$ is Boltzman's constant and $T$ is the temperature.   A one dimensional electron system in the energy window  $E_s \ll k_B T \ll E_c$ is said to be in the spin-incoherent regime.  If the charge sector is gapless it will be described by a zero-temperature Luttinger liquid theory and the resulting system is called a spin-incoherent Luttinger liquid (SILL).  Note that if $E_s$ refers to a spin {\em gap} then a SILL is still realized when    
$E_s \ll k_B T \ll E_c$ provided the charge sector remains gapless.  For a general discussion of the SILL see \cite{Fiete:rmp07}.  For SILL physics in ferromagnetic and superconducting hybrid structures see \cite{Tilahun:prb08,Tilahun:prb09}, and in cold atomic gases see \cite{Kakashvili:pra08}.

\section{Familar singular responses to a local time-dependent potential}

In this article we will only discuss a particularly simple form of a time-dependent spatially local potential whereby its value is changed instantaneously from zero to non-zero.  The prototypical physical realization of this situation is when a photon in the X-ray energy range ejects a deep ``core" electron from a metal--instantly changing the local charge by $+e$.  The electrons of the metal then rush in to screen the positive potential and thereby change their states.  If the electrons interact strongly with each other the processes of the electrons ``rushing in" to screen the local potential will obviously be impacted.  According to Luttinger liquid theory, interactions qualitatively change the nature of the low energy physics in such a way that electron-like quasi-particles are absent from the theory \cite{Voit:rpp95} leaving only the collective bosonic excitations.  On this basis alone, we expect the response of one-dimensional systems to differ in some ways from their higher dimensional counter parts.  In fact, this expectation is correct but to fully appreciate the differences, it is important to review some of basic ideas and qualitative expectations for the singular responses of non-interacting electron systems.

\subsection{Singular response of a non-interacting Fermi sea}

\begin{figure}
\begin{center}
\includegraphics[width=.8\linewidth]{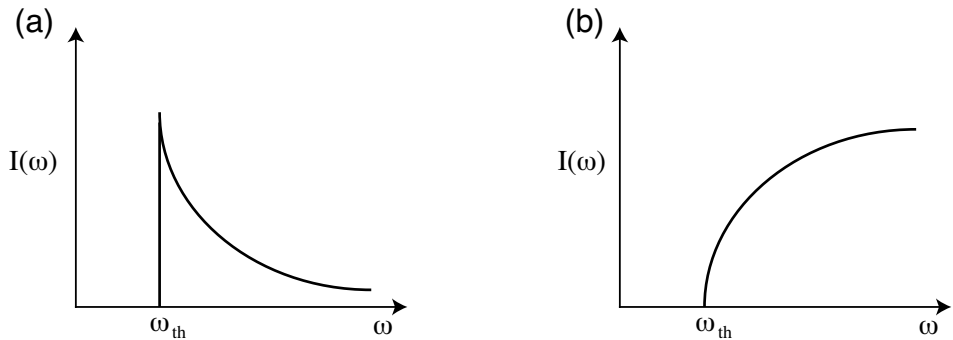} 
 \caption{\label{fig:schematic_threshold} Schematic behavior of photon absorption intensity, Eq.(\ref{eq:I_FL}), for an infinite mass core hole in three dimensions. The exponent $\alpha$ given by Eq.(\ref{eq:a}) contains positive contributions from excitonic effects and negative contributions from Anderson's orthogonality catastrophe.  It is possible for $\alpha$ to be either positive (a) or negative (b). In the spin-incoherent regime of one dimensional systems the absorption is not generally given by  Eq.(\ref{eq:I_FL}) but instead exhibits logarithmic corrections and additional physics enters the exponent $\alpha$.}
\end{center}
\end{figure}

\begin{figure}
\begin{center}
\includegraphics[width=.55\linewidth]{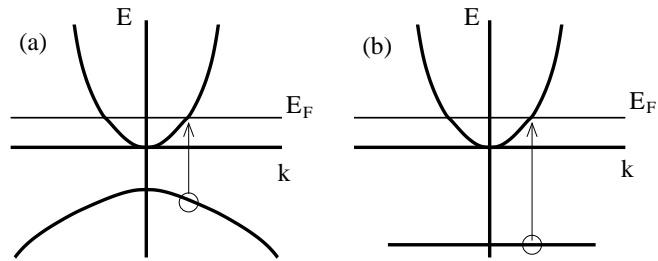} 
 \caption{\label{fig:direct_transition} Schematic of the possible direct transitions for exciting a hole with (a) finite mass and (b) infinite mass.  Here $E_F$ denotes the Fermi energy. In (a) an electron is excited from a valence band up to a conduction band leaving behind a single mobile hole in the valence band.  In (b) an electron is excited from an infinite mass core level.  The photo-absorption of both process (a) and (b) differ in one dimension compared to their counterparts in three dimensions.  Furthermore, the response in one dimension depends on whether the system is in the Luttinger liquid or spin-incoherent Luttinger liquid regime.}
\end{center}
\end{figure}

The seminal theoretical ideas of the singular response of a non-interacting electron gas due to a localized core hole were worked out by \cite{Mahan:pr67} and \cite{Nozieres:pr69}.  A nice textbook-level discussion can be found in \cite{Mahan} and a more recent review in \cite{Ohtaka:rmp90}.  The most important results to emerge are: (i) There is a threshold frequency $\omega_{\rm th}$ above which there is photon absorption and below which there is not. (ii) The  photon absorption at frequencies just above $\omega_{\rm th}$ follows a power law with exponent $\alpha$.  The photon absorption intensity is computed from a Fermi's Golden rule-type treatment in the electron-photon coupling and from the results just cited above takes the form
\begin{equation}
\label{eq:I_FL}
I(\omega)=A_0  \left(\frac{\omega_{\rm th}}{\omega-\omega_{\rm th}}\right)^\alpha\Theta(\omega-\omega_{\rm th}),
\end{equation}
near  threshold where $A_0$ is a constant, and $\Theta(x)=0$ for $x<0$ while  $\Theta(x)=1$ for $x>0$. 
A schematic of the possible behaviors is shown in Fig.~\ref{fig:schematic_threshold}. 
The important microscopic details of the electron dynamics and the electron-core hole interaction are encoded in the exponent $\alpha$ and the power-law form itself.  The power-law arises from summing an infinite series of diagrams containing logarithmic divergences \cite{Mahan:pr67}. As we will see, in the SILL the power law form is not sacred. It will be modified by logarithmic corrections, and the physics encoded in $\alpha$ will depend on whether or not the one-dimensional system is in the spin-incoherent regime. For a non-interacting electron gas the exponent takes the form \cite{Nozieres:pr69}
\begin{equation}
\label{eq:a}
\alpha=\frac{2\delta_0(k_F)}{\pi}-2\sum_l(2l+1)\left[\frac{\delta_l(k_F)}{\pi}\right]^2,
\end{equation}
where $\delta_l(k_F)$ is the phase shift induced by the core hole in the $l^{th}$ channel of the conduction electrons.  Often $s$-wave scattering ($l=0$) is the dominant channel and we will assume this is the case throughout this review. As the total amount of screening charge around the core hole must be one unit of charge, there is a Friedel sum rule:
\begin{equation}
1=\frac{2}{\pi}\sum_l (2l+1)\delta_l(k_F).
\end{equation}
If the scattering is not too strong, the phase shifts $\delta_l(k_F)$ can be computed in the Born approximation and one finds \cite{Mahan} $\delta_0(k_F)/\pi=N(E_F)V_0$ where $N(E_F)$ is the density of states at the Fermi energy and $V_0>0$ is the strength of the electron-hole interaction assumed momentum and energy independent.  (The actual core hole-electron interaction is $-V_0$ and therefore attractive.) Within the Born approximation and assuming only $s$-wave scattering, the exponent becomes
\begin{equation}
\label{eq:a_approx}
\alpha \approx 2 N(E_F)V_0-2[N(E_F)V_0]^2.
\end{equation}
Evidently, Eq.(\ref{eq:a}) and Eq.(\ref{eq:a_approx}) imply there are competing effects in $\alpha$.  The first term proportional to $\delta_0(k_F)$ is positive while the second term proportional to $[\delta_0(k_F)]^2$ is negative.  Going back to Eq.(\ref{eq:I_FL}) one sees that a positive contribution to $\alpha$ tends to make the threshold response more singular, while a negative contribution to $\alpha$ tends to make the threshold response less singular.  In fact, the origin of these two contributions are very well understood.  The first contribution to $\alpha$ in Eq.(\ref{eq:a}) comes from excitonic effects arising from the attractive interaction between the electrons and the core hole.  The second contribution to $\alpha$ in Eq.(\ref{eq:a}) arises from Anderson's orthogonality catastrophe \cite{Anderson:prl67} and it tends to suppress the absorption by the small (vanishing as $\omega \to 0$) overlap between initial and final states.  Note that for a non-interacting Fermi sea both excitonic effects and orthogonality effects disappear if $V_0=0$ and the absorption threshold becomes a simple step.   Looking ahead for a moment to Luttinger liquid physics, it is instructive to point out that if a collection of fermions is interacting, there will be orthogonality effects if an additional fermion of like charge is added to the system because the other fermions must adjust their state to accommodate the newcomer, but no excitonic effects.  This rearrangement of electronic states is what is responsible for the famous power law suppression of the tunneling density of states in a Luttinger liquid \cite{Giamarchi,Gogolin}.  Finally, it is worth emphasizing that depending on the microscopic details involved, it is possible for $\alpha$ to be positive or negative as shown in Fig.~\ref{fig:schematic_threshold}.  If the mass of the core hole is finite and one considers only direct transitions like those in Fig.~\ref{fig:direct_transition} one finds that the threshold is smeared out and disappears \cite{Hartmann:prb71,Ruckenstein:prb87}.  We will find this is not the case when we go to one dimension.

\subsection{Singular response of a Luttinger liquid}

Before we turn our attention to the spin-incoherent Luttinger liquid, it is useful for orientation to discuss the singular response of its closest cousin, the Luttinger liquid.  The behavior exhibited by the one-dimensional Luttinger liquid (LL) will already take us part of the way to understanding the response of the SILL.  In all of the results quoted below finite system size effects will not become important until frequencies smaller than $v_F/L$ are probed, where $L$ is the length of the system and $v_F$ is the Fermi velocity.  Also, finite temperature will cut-off the singular behaviors when relevant frequencies are of order $k_B T$ and smaller.  In this sense, finite system size and temperature can act as a non-zero threshold for tunneling into a finite length system or an infinite system at finite temperature, for example.

\subsubsection{Tunneling into a Luttinger liquid.}
\begin{figure}
\begin{center}
\includegraphics[width=.8\linewidth]{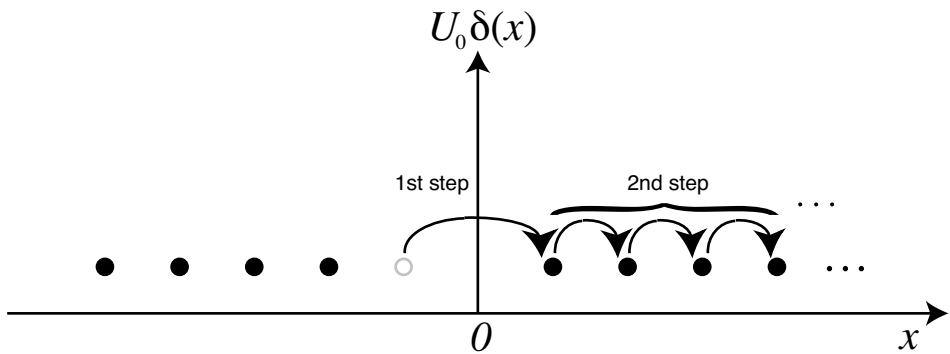} 
 \caption{\label{fig:Wigner_schematic} Schematic of the tunneling process of spinless electrons into a ``Wigner crystal" type electron arrangement. Electrons are schematically represented by black dots evenly spaced on either side of the tunneling barrier of strength $U_0$ at $x=0$. The tunneling occurs in two steps:  a fast process where the electron enters the region $x>0$ from the left followed by a slow process of charge relaxation where each electron shifts over to the right by a distance of one lattice constant.  At long times 
 $\langle \psi(\tau)\psi^\dagger(0)\rangle \sim \left( \frac{1}{\omega_D \tau} \right)^{1/g}$ as discussed in the text. The resulting behavior of the tunneling density of states is $A(\omega)\sim \omega^{1/g-1}.$  The qualitative behavior of the tunneling density of states looks very much like that in Fig.~\ref{fig:schematic_threshold}(b) with $\omega_{\rm th}=0$.}
\end{center}
\end{figure}

Let us first consider the simplest kind of singular response: abruptly adding a particle to a LL as occurs during tunneling.  A nice pedagogical discussion of the essential physics of tunneling into a LL is given by \cite{Fisher_Glazman}.  Here we adopt the central elements of their discussion.  The tunneling process is conveniently visualized within a Wigner crystal-type picture for spinless electrons, as shown in Fig.~\ref{fig:Wigner_schematic}.   
Such a picture implies very strong electron interactions, but the physical result and the final mathematical expression that describes it are actually valid at any interaction strength.  We imagine a process where an electron tunnels from the left of the barrier at $x=0$ to the right of the barrier at time $t=0$.  The rate at which the electron tunnels is determined by the Debye frequency of the Wigner crystal, $\omega_D =k_Fv$, where $k_F$ is the Fermi wave vector and $v$ is the phonon (charge) velocity.  When an electron tunnels the potential near the barrier suddenly (on a time scale $\sim 1/\omega_D$) changes.  The energy cost of this new configuration is large, $\sim \omega_D/g$, where $g \ll 1$ is the interaction parameter for the spinless Luttinger liquid.  (Here $g=1$ if the electrons are non-interacting.) This large energy cost is reduced by the relatively slow process of the electrons relaxing to new equilibrium positions in order to minimize the energy.  Once the relaxation process is complete, all the electrons will be shifted to the right by one crystalline period.  In the thermodynamic limit this new wavefunction is orthogonal to the original one which implies at zero energy ($\omega=0$) there is no tunneling.  This is a concrete example of Anderson's orthogonality catastrophe that we earlier mentioned plays a role in the exponent $\alpha$ of the threshold behavior of photon absorption.  For finite energy the final state may be in an excited state so the relaxation is not complete and the initial and final wavefunction can have finite overlap.  We will now see how this directly leads to a power law tunneling density of states, one of the ``singular" responses of interest in this article.

We can compute the tunneling density of states from the single particle Green's function \cite{Mahan}.  In the semi-classical approximation the electron tunneling takes the path of minimal action which is given by the classical trajectory
\begin{equation}
\langle \psi(x,t)\psi^\dagger(x,0)\rangle \sim \exp[-S(t)],
\end{equation}
where $\psi(x,t)$ is the electron annihilation operator at position $x$ and time $t$, and $\psi^\dagger(x,t)$ is the creation operator.  The initial density deformation $\rho(x,t=0)=2\delta(x)$ will eventually relax to $\rho(x,t=\infty)=0$.  As the processes is classically forbidden, the trajectory runs in imaginary time $t=i\tau$.  The classical equation of motions for a harmonic lattice give
\begin{equation}
\rho(x,\tau)=\frac{2}{\pi}\frac{v\tau}{x^2+(v\tau)^2}, \;\; x>0.
\end{equation}
The potential energy of the deformation is 
\begin{equation}
V_{\rm def}(\tau)=\frac{v}{g}\frac{\pi}{2}\int_0^\infty dx \rho(x,\tau)^2=\frac{2v}{\pi g}\int_0^\infty dx\frac{(v\tau)^2}{[x^2+(v\tau)^2]^2},
\end{equation}
which gives $V_{\rm def}(\tau)=1/(2g\tau)$.  By the virial theorem, the kinetic energy is equal to the potential energy. Therefore the action describing the slow part of the tunneling is
\begin{equation}
S(\tau)=2\int_{\omega_D^{-1}}^\tau d\tau' V_{\rm def}(\tau')=\frac{1}{g}\ln(\omega_D \tau),
\end{equation}
whence $\langle \psi(x,t)\psi^\dagger(x,0)\rangle \sim 1/(\omega_D \tau)^{1/g}$.  Taking the Fourier transform one finds the tunneling density of states $A(\omega)\sim \omega^{1/g-1}$. Since we have assumed $g<1$, the tunneling density of states $\to 0$ as $\omega \to 0$, as we noted above in the context of the orthogonality catastrophe.  

The discussion above illustrates some of the most central aspects of singular responses of interacting one-dimensional systems.  However, there is much more to the story.   Let us now suppose that instead of an electron tunneling, we consider an incident photon that abruptly creates a particle-hole pair as shown in Fig.~\ref{fig:direct_transition}. Note that due to the presence of the gap between the hole states and conduction band, there will be a minimum frequency $\omega_{\rm th}$ required to create the particle-hole pair.  If we first focus on the electrons in the conduction band, we see that the process is similar to tunneling: conduction band electrons see the sudden appearance of a new electron and they must adjust their positions by relaxing to a new state just as if the electron tunneled from outside the system.  As we have seen, this relaxation rate is related to the strength of the electron interactions $g$ in the Luttinger liquid, and based on the power law density of states we obtained, we expect the threshold behavior will also contain at least a power law contribution with an exponent that contains information about the interactions in the system.

It is also important to consider the interaction between the hole created and the electrons in the conduction band.  As holes and electrons generally have different masses and they attract rather than repel each other, the creation of a hole adds new physics to the electron tunneling physics above.  In fact, to compute the photon absorption spectrum near threshold we must {\em simultaneously} account for the interaction of the electrons with each other and with the hole. In general, this is a very difficult problem.  Fortunately, though, the power of Luttinger liquid theory and the commonly used technical methods such as bosonization allow us to make some precise and quite general statements about $I(\omega)$. 

\subsubsection{Fermi-edge singularity in a Luttinger liquid.}

Pioneering theoretical work in the study of the Fermi-edge singularity in Luttinger liquids was carried out more than 15 years ago \cite{Ogawa:prl92,Lee:prl92},  and later works soon followed \cite{Kane:prb94,Affleck:jpa94,Otani:prb96,Otani_2:prb96,Furusaki:prb97,Komnik:prb97,Tsukamoto:prb98,Tsukamoto:epj98,Balents:prb00}.  A recent review of the optical response of low dimensional systems (including some aspects of one-dimension) exists \cite{Ogawa:jpcm04}, so here we will only highlight the main issues and results to set the stage for our discussion of the spin-incoherent one-dimensional systems.

In one-dimensional electronic systems the effect of impurities is especially dramatic compared to three dimensions.  Even for an unrealistic non-interacting one-dimensional system the electrons localize leading to insulating behavior.  However, for a realistic system which contains repulsive interactions even a single local potential of arbitrary strength will localize the electrons to the half-line \cite{Kane:prl92,Furusaki:prb93}.  The central physics of this result is the relevance of electron backscattering as a perturbation: At low energies the strength of the impurity grows to arbitrarily large values and eventually divides the system into two semi-infinite segments.   This physics is important for the Fermi-edge singularity involving an infinitely massive core hole, {\it i.e.} one that is localized in space.

On the other hand, when the mass of the core hole is finite the physics is very different.  The properties of a heavy particle (holes are typically heavier than electrons) moving in a Luttinger liquid have been investigated \cite{Neto:prb96}.  It was shown that the mobility of the heavy particle diverges at low energies, implying that electron backscattering from it becomes irrelevant.  We therefore see that the Fermi-edge singularity in a Luttinger liquid falls into two categories: (i) infinite hole mass and (ii) finite hole mass.  Note that in three dimensions it was also important to divide the physics into the same two categories, but the reason for that division was different.   In three dimensions finite hole mass led to excitonic effects that smeared the edge singularity of the direct transition and eliminated the threshold. This has to do with a large (compared to one-dimension) phase space for recoil effects.  A sharp threshold only exists for the infinitely massive hole.  On the other hand, in one dimension the issue is about the relevance/irrelevance of backscattering and that is decided by the interactions (repulsive or attractive) in the LL.  Another way to see that it is natural to divide the classes into infinite and finite masses cases is that the symmetry is different.  An infinitely massive hole breaks the translational symmetry of the problem, while a hole with finite mass does not.

One of the first issues addressed regarding the Fermi-edge singularity in the LL is whether the threshold  still exists for the case of finite hole mass.  This was answered in the affirmative  \cite{Ogawa:prl92}. The next logical question is how does the threshold exponent $\alpha$ depend on the electron-hole and electron-electron interactions?  Initially it was argued that $\alpha$ was independent of the hole mass \cite{Ogawa:prl92}, but this was later challenged and corrected \cite{Castella:prb96,Tsukamoto:prb98,Tsukamoto:epj98}.  Knowing that any finite mass hole is an irrelevant perturbation \cite{Neto:prb96} we can immediately see that only forward scattering interactions will contribute to $\alpha$.  Indeed this is found as well as a schematic phase diagram for when $\alpha$ is positive or negative \cite{Ogawa:prl92}. The result is \cite{Tsukamoto:prb98}
\begin{equation}
\label{eq:a_LL_mobile}
\alpha^{LL}_{\rm mobile}=1-\frac{1}{2}\left[g(1-\delta_a)^2+\frac{1}{g}(1-\delta_s)^2\right],
\end{equation}
where $\delta_a$ is an asymmetric phase shift and $\delta_s$ a symmetric phase shift defined below in the discussion of the spin-incoherent case. The presence of $\delta_a$ implies that the exponent depends on the mass of the hole. Note that both $\delta_a$ and $\delta_s$ correspond to forward scattering processes, and  $g<1$ is the Luttinger parameter for repulsively interacting spinless electrons.  We can directly compare the result Eq.(\ref{eq:a_LL_mobile}) with the non-interacting case Eq.(\ref{eq:a}) by taking $g=1$ and $\delta_a=0$ (appropriate for an infinitely massive hole).  Doing so, we note the form is identical to an s-wave only scattering model.  From this comparison we also see that $\delta_s$ has the meaning of a phase shift up to factors of 2 and $\pi$.  Note that we are allowed to compare the result for a mobile impurity in a LL with an infinitely massive impurity in a free electron gas in three dimensions because in the latter backscattering is not a relevant perturbation.  They are therefore in a similar ``class".

The situation is different when we turn to immobile infinite mass holes.  Here, as before, a threshold exists, but now $\alpha$ is modified by the relevant backscattering and in fact obtains a universal contribution of 1/8 \cite{Kane:prb94,Prokofev:prb94,Affleck:jpa94,Furusaki:prb97,Komnik:prb97},  
\begin{equation}
\alpha^{LL}_{\rm immobile}=1-\frac{1}{g}(1-\delta_s)^2-1/8.
\end{equation}
In the next section we will discuss how these two results are modified in the spin-incoherent regime.  But before we do that, it is important to comment on the experimental situation.  Unfortunately, there are few published results \cite{Calleja:ssc91,Fritze:prb93,Calleja:prb95,Ihara:prl07} and the overall level of agreement between theory and experiment is poor.  Clearly there is great need for further experiments on Fermi-edge singularities on clean, single mode one-dimensional systems.  We hope this article will help to stimulate more.

\section{Tunneling and edge singularities in a spin-incoherent Luttinger liquid}

We finally come to the main topic of this review well prepared from the earlier discussions of singular responses of non-interacting electron gases and Luttinger liquids.  The question to ask is what are the similarities and differences between the singular responses of a SILL and a LL or Fermi liquid?  As we will see, the behavior is qualitatively different from both the LL and Fermi liquid making it possible to distinguish them in experiment.  Perhaps the simplest and most striking difference is the behavior of the tunneling density of states. We begin there.

\subsection{Tunneling into a spin-incoherent Luttinger liquid}

The first theoretical study of tunneling into a SILL was reported in the infinite $U$ limit of the Hubbard model in a lengthy Bethe ansatz calculation \cite{cheianov03,Cheianov04}.  Immediately afterwards a simpler and more general method based on bosonization and the imaginary time path integral representation was given \cite{Fiete:prl04}.  Extensions of the method to finite length systems including boundary effects and externally applied magnetic fields have also been reported \cite{Fiete:prb05,Kindermann_crossover:prb06,Kakashvili:prb07,Fiete:rmp07}.  While the imaginary time path integral formulation is well suited to the regime of the SILL, it does not provide a conceptually straightforward generalization to the regime $k_B T \lesssim E_s \ll E_c$.  In order to address that regime, we need a representation of the spin degrees of freedom that are valid at arbitrary energy scales. Such a formulation is now in hand \cite{Matveev:prl07,Matveev:prb07} and it provides a good starting point for numerical studies in the regime $k_B T \lesssim E_s \ll E_c$.

\subsubsection{Bosonization of strongly interacting electrons.}
As is typical of interacting one dimensional electron systems, we assume the Hamiltonian in the strongly interacting case is spin-charge
separated \cite{Matveev:prb04,Fiete:rmp07}, $H=H_c+H_s$.  Here, $H_c$ is identical to that of the LL, and is given by \cite{Giamarchi,Gogolin}
\begin{equation}
H_c  = v_c\int \frac{dx}{2\pi} \left[\frac{1}{g}(\partial_x \theta(x))^2 + g(\partial_x \phi(x))^2\right],
\label{eq:H_holon}
\end{equation}
where $\theta$ and $\phi$ are bosonic fields satisfying $[\theta(x),\phi(x')]=-i\frac{\pi}{2} {\rm sgn}(x-x')$, and $g=2g_c$ is the effective Luttinger interaction parameter of the electron gas \cite{Fiete_2:prb05}. 
 
On the other hand, the spin Hamiltonian at arbitrary temperatures is to a very good approximation given by a nearest neighbor antiferromagnetic Heisenberg spin chain \cite{Matveev:prb04,Klironomos:prb05,Fogler_exch:prb05},
\begin{equation}
\label{eq:S_chain}
H_s = \sum_l J {\vec S}_l\cdot {\vec S}_{l+1},
\end{equation}
where evidently the spin energy is set by $J$: $E_s=J$.  The basic idea is to represent the electron operator as a product of operators that describe the holons $\Psi(x)$ [spinless fermions that naturally arise in the context of strongly interacting fermions and the
spin-incoherent regime \cite{Fiete_2:prb05,Fiete:prl04}] and the spin degrees of freedom ${\vec S_l}$.  The holon operators (denoted by
$\Psi^\dagger,\Psi$) by construction satisfy the equation
\begin{equation}
\label{eq:Psi}
\Psi^\dagger(x)\Psi(x)=\psi_\uparrow^\dagger(x)\psi_\uparrow(x)+\psi_\downarrow^\dagger(x)\psi_\downarrow(x),
\end{equation}
where $\psi_s$ is the electron annihilation operator for electrons of spin projection $\sigma$, and $\psi^\dagger_\sigma$ the corresponding electron
creation operator.

The issue of how to bosonize the electron operator for a strongly interacting system earlier arose in the context of the large $U$ limit of the
one-dimensional Hubbard model where \cite{Penc:prl95} wrote the electron creation operator as
\begin{equation}
\label{eq:Penc_bos}
\psi_\sigma^\dagger(0)=Z^\dagger_{0,\sigma} \Psi^\dagger(0),
\end{equation}
where $Z^\dagger_{0,\sigma}$ creates a site on the spin chain Eq.(\ref{eq:S_chain}) with spin projection $\sigma$.  The expression in Eq.(\ref{eq:Penc_bos}) can be
physically motivated as follows.  From Eq.(\ref{eq:Psi}) it is clear that the creation of an electron is also accompanied by the creation of a
holon.  However, electrons also carry spin so there must be a component of the electron operator that also creates spin.  This is accomplished
by $Z^\dagger_{0,\sigma}$.  In general, one has $Z^\dagger_{l,\sigma}$ as the object that adds a new site to the spin chain between $l-1$ and $l$.  While
this appears physically intuitive, the expression suffers from the drawback that it does not naturally account for the variation of electron
density with position in a real electron gas \cite{Penc:prl96}.  The remedy for this
issue \cite{Matveev:prb07}  is to define the position at which the spin site is added to the chain Eq.(\ref{eq:S_chain}) in terms of the number of holons to the left of the site,
\begin{equation}
\label{eq:l_x}
l(x)=\int_{-\infty}^x \Psi^\dagger(y)\Psi(y) dy.
\end{equation}
In terms of Eq.(\ref{eq:l_x}) the electron creation and annihilation operators are defined as
\begin{eqnarray}
\label{eq:psi+_x}
\psi_\sigma^\dagger(x)&=&Z^\dagger_{l(x),\sigma} \Psi^\dagger(x),\\
\label{eq:psi_x}
\psi_\sigma(x)&=&\Psi(x) Z_{l(x),\sigma}.
\end{eqnarray}
The operators given above explicitly account for the fact that the spins are attached to electrons, and the formulas are valid at all energy
scales. It is perhaps worth noting that even though the Hamiltonian is spin-charge separated, the electron operators are not written as a
product of a spin piece and a charge piece because the ``spin" pieces $Z_{l(x),\sigma}$ also depend on the electron density via Eq.(\ref{eq:l_x}).

In writing Eq.(\ref{eq:psi+_x}) and Eq.(\ref{eq:psi_x}) no assumptions have been made about the energy scale relative to the spin and charge energies. We now restrict our considerations to energies small compared to $E_c$, but arbitrary with respect to $E_s$.  In this case, we are free
to bosonize the holon sector.  We have already given the form of the charge Hamiltonian in Eq.(\ref{eq:H_holon}).  The spinless fields $\theta$ and $\phi$ can be related to the holon density as \cite{Fiete:prl04,Fiete:prb05}
\begin{equation}
\label{eq:den_rel}
\Psi^\dagger(x)\Psi(x)=\frac{1}{\pi}[k_F^h + \partial_x \theta(x)],
\end{equation}
where the holon Fermi wave vector is twice the electron Fermi wavevector \cite{Fiete:prl04} $k_F^h=2k_F$ and the bosonic fields satisfy the
commutation relations $[\theta(x),\partial_y\phi(y)]=i\pi\delta(x-y)$.

Since we are interested in low energies with respect to the charge energy, the electron operator may be expanded about the two holon Fermi
points at $\pm k_F^h$,
\begin{equation}
\label{eq:Psi_RL}
\Psi(x)=\Psi_R(x)+\Psi_L(x),
\end{equation}
where $\Psi_R(x)$ destroys an holon near the right Fermi point and $\Psi_L(x)$ destroys an electron near the left Fermi point. The left and
right holon operaters are bosonized as
\begin{equation}
\label{eq:Psi_RL_fields}
\Psi_{R,L}(x)=\frac{1}{\sqrt{ 2\pi \alpha_c}} e^{-i\phi(x)}e^{\pm i[k_F^h+\theta(x)]},
\end{equation}
where $\alpha_c$ is a short distance cut-off of order the interparticle spacing $a$.  Combining the results of Eq.(\ref{eq:l_x}), Eq.(\ref{eq:psi_x}),
Eq.(\ref{eq:den_rel}), Eq.(\ref{eq:Psi_RL}) and Eq.(\ref{eq:Psi_RL_fields}) one obtains the bosonized form of the electron annihilation operator for spin $\sigma$
\begin{eqnarray}
\label{eq:psi_bos}
\psi_\sigma(x)=\frac{e^{-i\phi(x)}}{\sqrt{2 \pi \alpha_c}}\left(e^{i[k_F^hx+\theta(x)]}+e^{-i[k_F^hx+\theta(x)]}\right) Z_{l,\sigma}\bigg |_{l=\frac{1}{\pi}[k_F^hx+\theta(x)]},
\end{eqnarray}
and an analogous expression for the electron creation operator $\psi^\dagger_\sigma(x)$.  Expression Eq.(\ref{eq:psi_bos}), however, it not quite
complete as it does not account for the discreteness of the charge of the electron.  This can be accomplished by interpreting
\begin{equation}
\label{eq:Z_proper}
Z_{l,\sigma}\bigg |_{l=\frac{1}{\pi}[k_F^hx+\theta(x)]} \to \sum_l Z_{l,\sigma}\delta\left({1 \over \pi}[k_F^hx+\theta(x)]-l\right),
\end{equation}
after which the full electron annihilation operator (include both left and right moving parts) becomes \cite{Matveev:prb07}
\begin{equation}
\label{eq:psi_final_bos}
\psi_\sigma(x)=\frac{e^{-i\phi(x)}}{\sqrt{2\pi\alpha_c}} \int_{-\infty}^\infty \frac{dq}{2\pi} z_\sigma(q) e^{i(1+{q \over \pi})[k_F^hx+\theta(x)]},
\end{equation}
where
\begin{equation}
\label{eq:little_z}
z_\sigma(q)=\sum_{l=-\infty}^\infty Z_{l,\sigma}e^{-iql}.
\end{equation}
The expression for the electron annihilation operator Eq.(\ref{eq:psi_final_bos}) is the key result of \cite{Matveev:prb07} who also showed that in the limit of small energies compared to $E_s$ the expression correctly recovers the standard LL formulas for the electron annihilation operator.  With Eq.(\ref{eq:psi_final_bos})  correlation functions involving
electron operators can be expressed in terms of the correlation functions of the holon and spin sectors at arbitrary temperatures with respect to $E_s$, but small energies compared to $E_c$. 

\subsubsection{Tunneling density of states in a SILL.}
The tunneling density of states in a SILL is qualitatively different from the LL.  While the tunneling density of states in a LL qualitatively resembles the behavior shown in Fig.\ref{fig:schematic_threshold}(b) with $\omega_{\rm th}=0$, the tunneling density of a SILL qualitatively resembles the behavior in   Fig.\ref{fig:schematic_threshold}(a) with $\omega_{\rm th}=0$, at least for a range of frequencies, $k_B T \lesssim \hbar \omega \ll E_c$ and for not too strong interactions.  For this reason, the tunneling density of states may provide the simplest and most transparent test for SILL physics in experiment.

Tunneling into an infinitely long SILL has been considered by \cite{cheianov03,Fiete:prl04,Matveev:prl07}.  Deep in the spin-incoherent regime, the first-quantized path integral representation of the bosonized electron's Green's function is a convenient method \cite{Fiete:prl04,Fiete:prb05,Fiete:rmp07}. Here we instead follow \cite{Matveev:prl07,Matveev:prb07} as it provides a better launching point for numerical studies in the regime $k_B T \lesssim E_s \ll E_c$, and also readily captures some other important features of the tunneling density of states.

The tunneling density of states, $\nu(\omega)=-\frac{1}{\pi}{\rm Im}[G^R(x=0,\omega)]$, where $G^R(x=0,\omega)$ is the Fourier transform of the retarded Green's function. The retarded Green's function generally contains particle and hole contributions, but for illustrative purpose, we will only focus on the particle contributions.  Hole contributions are conceptually and technical similar, and in the spin-incoherent regime it can be shown that $\nu(\omega)=2\nu(-\omega)$ for $\hbar \omega \gg k_B T$  \cite{Matveev:prl07}.  Here we focus on the particle contribution: $G^+_\sigma(x,t)=\langle
\psi_\sigma(x,t)\psi_\sigma^\dagger(0,0)\rangle$. Using the general formulas Eq.(\ref{eq:psi_final_bos}) and Eq.(\ref{eq:little_z}) we have
\begin{eqnarray}
G^+_\sigma(x,t)=\frac{1}{2\pi \alpha_c} \int_{-\infty}^{\infty} \frac{dq_1}{2\pi} \int_{-\infty}^{\infty} \frac{dq_2}{2\pi}
\sum_{l_1,l_2} e^{-i(q_1 l_1-q_2l_2)}  \nonumber \\
\times \langle Z_{l_1,\sigma} Z^\dagger_{l_2,\sigma} \rangle \langle e^{i[(1+\frac{q_1}{\pi})[k_F^hx+\theta(x,t)]-\phi(x,t)]}  e^{-i[(1+\frac{q_2}{\pi})\theta(0,0)-\phi(0,0)]}\rangle,\;\;\;
\end{eqnarray}
where we have assumed the time evolution of the spin degrees of freedom are slow compared to the charge degrees of freedom so that we can neglect their time dependence \cite{Matveev:prb07}.
In a translationally invariant system, one has $\langle Z_{l_1,\sigma} Z^\dagger_{l_2,\sigma} \rangle=\langle Z_{l_1-l_2,\sigma} Z^\dagger_{0,\sigma} \rangle$.  Making a change of variables $l=l_1-l_2$, the summation over the sites of the spin chain results in a delta function, $2\pi \delta(q_1-q_2)$, which immediately kills one of the momentum integrals and sets $q_1=q_2$. The resulting Green's function is
\begin{eqnarray}
\label{eq:G_final_trans}
G^+_\sigma(x,t)=\frac{1}{2\pi \alpha_c} \int_{-\infty}^{\infty} \frac{dq}{2\pi}c^+_\sigma(q)e^{i(1+\frac{q}{\pi})k_F^hx}g_q^+(x,t),\nonumber \\
 \end{eqnarray}
where 
\begin{equation}
\label{eq:c}
c^+_\sigma(q)=\sum_{l=-\infty}^\infty \langle Z_{l,\sigma} Z^\dagger_{0,\sigma} \rangle e^{-iql},
\end{equation}
 and
\begin{equation}
\label{eq:g}
g_q^+(x,t)= \langle e^{i[(1+\frac{q}{\pi})\theta(x,t)-\phi(x,t)]}  e^{-i[(1+\frac{q}{\pi})\theta(0,0)-\phi(0,0)]}\rangle.
\end{equation}
Since the holon sector is described the the Gaussian theory Eq.(\ref{eq:H_holon}) we can make use of the identity $\langle e^{i {\cal O}}\rangle= e^{-\langle {\cal O}^2\rangle/2}$ for operator $\cal O$ and evaluate the correlation function Eq.(\ref{eq:g}) at zero temperature with respect to the charge energy.  [Generalization to small but finite temperatures in the charge sector is trivial \cite{Giamarchi,Gogolin}.]  The expression for the hole-like contributions to the Green's function has the same structure as Eq.(\ref{eq:G_final_trans}) and is reported in \cite{Matveev:prb07}.  Since for low temperatures compared to the charge energy the holon correlations in Eq.(\ref{eq:g}) can be analytically evaluated in a straightforward way, the emphasis for arbitrary temperatures with the respect to the spin energy is on the spin correlations described by Eq.(\ref{eq:c}).   For arbitrary temperatures the evaluation of $c^+_\sigma(q)$ is non-trivial \cite{Matveev:prb07} and it is precisely at this point that numerical input for both zero and non-zero magnetic fields would be most helpful.  Keeping for the moment temperature arbitrary with respect to $E_s$, but effectively zero compared to $E_c$ one finds for the tunneling density of states \cite{Matveev:prl07},
\begin{equation}
\label{eq:tdos}
\nu(\omega)=\nu_0\int \frac{dq}{2\pi} \frac{c^+_\sigma(q)}{\Gamma(\lambda(q)+1)}\left(\frac{\hbar \omega}{E_c}\right)^{\lambda(q)},
\end{equation}
where $\nu_0=(\pi\hbar v_c)^{-1},\;\lambda(q)=\frac{1}{2}\left [ g(1+\frac{q}{\pi})^2+\frac{1}{g}\right]-1$ and $g=2g_c$ is the Luttinger interaction parameter appearing in Eq.(\ref{eq:H_holon}).  From the exponent $\lambda(q)$ one sees that the frequency dependence $\sim \left(\frac{\hbar \omega}{E_c}\right)^{1/(2g)-1}e^{-\ln(E_c/\hbar \omega)(1+q/\pi)^2/2}$ which is maximum for $q=-\pi$.  Expanding about this point and doing the Guassian integration, one finds
\begin{equation}
\label{eq:nu_structure}
\nu(\omega)\approx  \nu_0 \frac{c^+_\sigma(-\pi)}{2\pi \Gamma(\lambda(-\pi)+1)}\left(\frac{\hbar \omega}{E_c}\right)^{1/(2g)-1}\sqrt{\frac{2\pi}{g\ln(E_c/\hbar \omega)}},
\end{equation}
which agrees with the result earlier obtained for the infinite $U$ limit of the Hubbard model \cite{cheianov03} and for general interaction parameter $g$ in the spin-incoherent regime \cite{Fiete:prl04}.  Note, however, that the result Eq.(\ref{eq:nu_structure}) is valid at $T=0$.  The only assumptions were that the dominant weight comes from  the point $q=-\pi$ and that $c_\sigma^+(-\pi)\neq 0$. 
If one is deep in the spin-incoherent regime $\langle Z_{l,\sigma} Z^\dagger_{0,\sigma} \rangle = \left(\frac{1}{2}\right)^{|l|}$ {\em independent of the details and form} of $H_s$ [a simple argument for this is given in \cite{Fiete:prl04,Fiete:rmp07})], and one has $c^+_\sigma(q)=\frac{3}{5-4\cos(q)}$.  Hence, the assumptions leading to Eq.(\ref{eq:nu_structure}) are valid and the result indeed applies in the spin-incoherent regime as well.  On the other hand, if the spins are fully polarized (along the z-axis, say) then $c_\uparrow^+(q)=2\pi\delta(q)$ and the tunneling density of states will necessarily have a different form (in fact, the familiar LL form) from Eq.(\ref{eq:nu_structure}) \cite{Matveev:prl07}.   Also at $T=0$ and zero magnetic field one has $c_\sigma^+(-\pi)=0$ so the logarithmic factor appearing in  Eq.(\ref{eq:nu_structure})  is absent and the LL form is obtained.

Finally, note that in the spin-incoherent regime where Eq.(\ref{eq:nu_structure}) applies the tunneling response of a SILL looks qualitatively like the behavior in Fig.~\ref{fig:schematic_threshold}(a) with $\omega_{\rm th}=0$ for $g=2g_c>1/2$, but it {\em does not} follow the analytical form of Eq.(\ref{eq:I_FL}) because of the logarithmic corrections.  If the interactions are both strong and long-ranged, it is possible to obtain $g_c<1/4$ in which case the tunneling density of states would look qualitatively like the behavior in Fig.~\ref{fig:schematic_threshold}(b) with $\omega_{\rm th}=0$ although it would still not fit the analytic form in Eq.(\ref{eq:I_FL}).  [Recall that for zero-range interactions the smallest value $g_c$ can take is $1/2$, obtained in the infinite-$U$ limit of the Hubbard model \cite{Giamarchi}.]  In this sense, the tunneling response of the SILL is in a different universality class from both the Fermi liquid and Luttinger liquid.  We will now see that the same logarithmic corrections can also appear in the frequency dependence of the photon absorption.

\subsection{Fermi-edge singularity in a spin-incoherent Luttinger liquid} 

As we discussed earlier, the physics of the Fermi-edge singularity falls into two classes: (1) Electron backscattering from the hole is an irrelevant perturbation. (2) Electron backscattering from the hole is a relevant perturbation.  For a Luttinger liquid these two regimes are distinguishes by whether the hole has a finite mass, case (1), or an infinite mass, case (2).  In the SILL there is a subtlety because the relevance or irrelevance of backscattering from a local potential (for example, an infinite mass core hole) depends on the interactions in the system through the Luttinger parameter $g$.  In particular, a local impurity is only a relevant perturbation if $g_c<1/2$  [which is the same as $g<1$] \cite{Fiete_2:prb05}.  If  $1/2<g_c<1$ the interactions of LL are repulsive, but a local impurity is an {\em irrelevant} perturbation.  Therefore, in the SILL we effectively have 3 possible regimes for the Fermi-edge singularity: (i) Finite mass mobile hole for which backscattering is irrelevant.  (ii) Infinite mass core hole for which backscattering is irrelevant.  (iii) Infinite mass core hole for which backscattering is relevant. 

Let us first consider the case of a finite mass hole.  In this case the backscattering is always an irrelevant perturbation.  For addressing the threshold behavior of the photon absorption we can therefore neglect backscattering terms and focus only on forward scattering interactions between the electron and the hole.  In the initial study of the Fermi-edge singularity in a LL \cite{Ogawa:prl92} it was incorrectly concluded that the exponent $\alpha$ is independent of the dynamics of the hole, {\it i.e.} its mass.  A more careful treatment showed that indeed the mass of the hole enters $\alpha$ \cite{Tsukamoto:prb98,Tsukamoto:epj98}.  The work of Tsukamoto {\em et al.} employed a combined Bethe ansatz-Conformal Field Theory study of a zero range interaction between particles.  The asymptotic structure of the energy and critical exponents in this model could be obtained exactly.  From the formulas obtained it was shown explicitly that the exponent $\alpha$ depends on the mass of the hole.  Perhaps the most important result is the structure of the electron-hole interaction:  There are actually two forward scattering terms that emerge.  One is the familiar density-density interaction between hole density and electron density.  The other is less familiar and it describes the interaction of the hole with a current of electrons that is flowing by it.  In other words, it is natural to transform to frame co-moving with the hole \cite{Neto:prb96}.  In this frame the hole sees a current of electrons.  One might think that such a picture would break time-reversal symmetry, but in an actual experiment electrons would be excited near both the right and left Fermi points shown in Fig.\ref{fig:direct_transition}(a).  Thus, as a whole, time-reversal symmetry is preserved.  As the contribution to the absorption edge is the same in both cases, one need only focus on one Fermi point.  Our discussion below will cover both Fermi points in one fell swoop, as the electron operator Eq.(\ref{eq:psi_final_bos}) contains both right and left moving pieces.

The Fermi-edge singularity in a spin-incoherent Luttinger liquid was discussed by \cite{Fiete:prl06}.  The full Hamiltonian for our problem in the frame co-moving with the hole is $H=H_{\rm elec}+H_{\rm elec-hole}+H_{\rm hole}$, where $H_{\rm elec}$ is given by Eq.(\ref{eq:H_holon}) and Eq.(\ref{eq:S_chain}), and 
\begin{equation}
\label{eq:e-h}
H_{\rm elec-hole}=\frac{U_s^f}{\pi} h^\dagger h \partial_x\theta(0) \pm  \frac{U_a^f}{\pi}h^\dagger h \partial_x\phi(0)\,,
\end{equation}
with  $H_{\rm hole}=\sum_\sigma E_{h,\sigma} h_\sigma^\dagger h_\sigma$ and $h^\dagger h=\sum_\sigma h_\sigma^\dagger h_\sigma$.  Here $U_s^f$ is the symmetric part of the forward scattering from the hole and $U_a^f$ is the antisymmetric part of the forward scattering \cite{Tsukamoto:prb98}. 
(In our convention $\partial_x\theta$ represents the density fluctuations and $\partial_x\phi$ the particle current.)  Again, the antisymmetric part appears since in the frame of the hole, it sees a net current of particles scattering from it. The ``+'' sign is for a right-moving hole and the ``-'' sign is for a left-moving hole. The parameter $U_a^f$ depends on the momentum and mass of the hole, and when it is at rest,  $U_a^f\equiv 0$ \cite{Tsukamoto:prb98}. The operator $h_\sigma^\dagger$ creates a hole with spin $\sigma$ and $h_\sigma$ annihilates a hole with spin $\sigma$.  The energy of such a hole in its rest frame is $E_{h,\sigma}$, which in the presence of an externally applied magnetic field can depend on $\sigma$.
 
To compute the photon absorption we treat the electron-photon interaction in second order perturbation theory.  Standard manipulations show \cite{Mahan}
\begin{equation} 
\label{eq:threshold}
I(\omega) \propto \sum_\sigma {\rm Re}\int_0^\infty dt e^{i\omega t} \langle \psi_\sigma(t)h_\sigma(t)h_\sigma^\dagger(0)\psi_\sigma^\dagger(0)\rangle.
\end{equation}
To evaluate the correlations appearing in Eq.(\ref{eq:threshold}) we must diagonalize the Hamiltonian.
The Hamiltonian $H$ can be diagonalized with the unitary transformation
\begin{equation}
U={\rm exp}\{-i[\delta_a\theta(0)+\delta_s\phi(0)]h^\dagger h\}
\end{equation}
 where 
 \begin{equation}
 \delta_a\equiv \mp U_a^f/(vg\pi),
 \end{equation}
  and 
 \begin{equation}
 \delta_s \equiv -gU_s^f/(v\pi).
 \end{equation}
Applying this transformation we find $\bar H  \equiv U^\dagger H U =  H_{\rm elec}+\bar H_{\rm hole}$, where the only change to $H_{\rm hole}$ is a shift in the hole energy $E_{h,\sigma} \to \tilde E_{h,\sigma}$, which is unimportant to us here.  Our main concern is with computing the threshold exponent $\alpha$ and the functional form of the response. 

We begin with the evaluation of the correlation function $C_\sigma(\tau) =\langle \psi_\sigma(\tau)h_\sigma(\tau)h_\sigma^\dagger(0)\psi_\sigma^\dagger(0)\rangle$, where $\tau$ is the imaginary time.  At finite temperature, 
\begin{eqnarray}
C_\sigma(\tau)&=&\frac{1}{Z}{\rm Tr}\left[e^{-\beta H}  \psi_\sigma(\tau)h_\sigma(\tau)h_\sigma^\dagger(0)\psi_\sigma^\dagger(0)\right] \\
&=&\frac{1}{Z}{\rm Tr}\left[e^{-\beta \bar H}  \bar \psi_\sigma(\tau)\bar h_\sigma(\tau)\bar h_\sigma^\dagger(0)\bar \psi_\sigma^\dagger(0)\right],
\end{eqnarray}
where in the second line we have introduced the unitary transformation $U^\dagger U=1$ and used the cyclic property of the trace. We have already determined $\bar H$, so it remains to determine $\bar \psi_\sigma$ and $\bar h_\sigma$.   Direct evaluation gives $\bar \psi_\sigma = \psi_\sigma$ (up to unimportant multiplicative factors) and $\bar h_\sigma=h_\sigma e^{-i[\delta_a\theta+\delta_s\phi]} $.  Therefore, the correlation function $C_\sigma(\tau)$ separates as $C_\sigma(\tau) = C_{\psi,\sigma}(\tau)C_{h,\sigma}(\tau)$.  One readily finds $C_{h,\sigma}(\tau)=e^{-\tilde E_{h,\sigma} \tau}$ at zero temperature.  This factor will enter the threshold frequency $\omega_{\rm th}$ in Eq.(\ref{eq:threshold}) and will in general also depend on the external magnetic field.  Our main interest here is in the evaluation of the part of the correlation function that will give us the frequency dependence just above threshold,
\begin{eqnarray}
\label{eq:C_psi}
C_{\psi,\sigma}(\tau)=\frac{1}{Z_{\rm elec}}{\rm Tr}\biggl[e^{-\beta H_{\rm elec}}   \psi_\sigma(\tau) e^{-i[\delta_a\theta(\tau)+\delta_s\phi(\tau)]}  e^{i[\delta_a\theta(0)+\delta_s\phi(0)]} \psi_\sigma^\dagger(0)\biggr],
\end{eqnarray}
where $Z_{\rm elec}={\rm Tr}[e^{-\beta H_{\rm elec}}]$.  Formally, Eq.(\ref{eq:C_psi}) bears a striking resemblance to the single particle Green's function evaluated in the spin-incoherent regime above.  It is worth pausing a moment to understand the physics of Eq.(\ref{eq:C_psi}).  Recalling that the operator $e^{i \phi(x,\tau)}$ creates a particle at space-time point $(x,\tau)$ in the many-body system and the electron number is related to the $\theta$ field via $N(x,\tau)=\bar nx+\frac{1}{\pi}(\theta(x,\tau)-\theta(0,0))$, with $\bar n$ the average particle density, we see immediately that $C_{\psi,_\sigma}(\tau)$ involves adding an electron {\em plus} an additional ``background excitation" (from $e^{i\delta_s \phi(0)}$) and then removing  the same particle and its additional background at a time $\tau$ later. The factors $e^{\pm i \delta_a \theta}$ contribute additional density fluctuations coming from the motion of the finite mass valence hole.  This factor is absent in the infinite mass limit. 

It remains to evaluate the trace in Eq.(\ref{eq:C_psi}).  We again follow the more general bosonization scheme for strongly interacting systems described above which reduces the to previously obtained results deep in the spin-incoherent regime \cite{Fiete:prl06}.  We have
\begin{equation}
C_{\psi,\sigma}(\tau)=\frac{1}{2\pi \alpha_c} \int_{-\infty}^{\infty} \frac{dq}{2\pi}c^+_\sigma(q)f_q^+(\tau),
 \end{equation}
where
\begin{equation}
\label{eq:f}
f_q^+(\tau)= \langle e^{i[(1+\frac{q}{\pi}-\delta_a)\theta(\tau)-(1-\delta_s)\phi(\tau)]}  e^{-i[(1+\frac{q}{\pi}-\delta_a)\theta(0)-(1-\delta_s)\phi(0)]}\rangle.
\end{equation}
In the zero temperature limit with respect to the charge degrees of freedom, we can evaluate Eq.(\ref{eq:f}) using Eq.(\ref{eq:H_holon}) as we did for the single-particle Green's function.  The result is
\begin{equation}
f_q^+(\tau)=\left(\frac{\alpha_c}{v_c \tau+\alpha_c}\right)^{\frac{1}{2}[g(1+\frac{q}{\pi}-\delta_a)^2+\frac{1}{g}(1-\delta_s)^2]}.
\end{equation}
The frequency dependence of $I(\omega)$ can now be computed from Eq.(\ref{eq:threshold}) using $C_\sigma(\tau) = C_{\psi,\sigma}(\tau)C_{h,\sigma}(\tau)$.  The computation is essentially identical to that of the tunneling density of states, Eq.(\ref{eq:tdos}), only $\omega \to \tilde \omega= \omega-\omega_{\rm th}$ and $\lambda(q) \to \tilde \lambda(q)=\frac{1}{2}[g(1+\frac{q}{\pi}-\delta_a)^2+\frac{1}{g}(1-\delta_s)^2]-1$.  Therefore, we have
\begin{equation}
\label{eq:I_gen}
I(\omega) \propto \sum_\sigma \int \frac{dq}{2\pi} \frac{c^+_\sigma(q)}{\Gamma(\tilde \lambda(q)+1)}\left(\frac{\hbar \tilde \omega}{E_c}\right)^{\tilde \lambda(q)}\Theta(\tilde \omega),
\end{equation}
where the $\Theta(\tilde \omega=\omega-\omega_{\rm th})$ factor tells us there is no photon absorption below below a threshold frequency $\omega_{\rm th}$.  From Fig.(\ref{fig:direct_transition})(a) it is clear that $\hbar \omega_{\rm th}$ is of order the band gap plus the Fermi energy of the electrons plus the absolute value of the energy of the hole near the Fermi points of the electron gas.  As with the expression Eq.(\ref{eq:tdos}) for the tunneling density of states, Eq.(\ref{eq:I_gen}) is the general expression for the photon absorption in an interacting one-dimensionsal system.  For temperatures low compared to the charge energy, but arbitrary with respect to the spin energy, the heavy lifting is in the evaluation of $c^+_\sigma(q)$ as we emphasized in the discussion of the single-particle Green's function.  Following the same arguments we used there, we see that the frequency dependence $\sim \left(\frac{\hbar \tilde \omega}{E_c}\right)^{\frac{1}{2g}(1-\delta_s)^2-1}e^{-\frac{g}{2}\ln\left[\frac{E_c}{\hbar \tilde \omega}\right](1+q/\pi-\delta_a)^2}$ which is a maximum for $q=\pi(\delta_a-1)$.  Expanding about this point and doing the Guassian integration, one finds
\begin{equation}
\label{eq:I_mobile_final}
I(\omega) \propto \sum_\sigma \frac{c^+_\sigma(\pi(\delta_a-1))}{2\pi \Gamma(\tilde \lambda(\pi(\delta_a-1))+1)}\left(\frac{\hbar \tilde \omega }{E_c}\right)^{\frac{1}{2g}(1-\delta_s)^2-1}\sqrt{\frac{2\pi}{g\ln(\frac{E_c}{\hbar \tilde\omega})}}\Theta(\tilde \omega).
\end{equation}
Note that the final form of the photon absorption Eq.(\ref{eq:I_mobile_final}) is almost identical to that of the tunneling density of states, Eq.(\ref{eq:nu_structure}), aside from the factor $\delta_s$ that appears in the threshold exponent.

It is instructive to compare the similarities and differences of the Fermi-edge singularity with a mobile hole in the LL and SILL.  Comparing with the general formula Eq.(\ref{eq:I_FL}) we see that the threshold behavior is different in the SILL compared to the LL and even the Fermi liquid.  The threshold behavior of the SILL contains the same logarithmic corrections that we found earlier in the tunneling density of states.  Focusing on the power-law part of the threshold behavior we see that
\begin{equation}
\label{eq:a_SILL_mobile}
\alpha^{SILL}_{\rm mobile}=1-\frac{1}{2g}(1-\delta_s)^2,
\end{equation}
so that it is independent of $\delta_a$, unlike the LL case in Eq.(\ref{eq:a_LL_mobile}).  Since the asymmetric phase shift does not appear, this implies the threshold exponent is {\em independent of the hole dynamics and therefore the mass of the hole} in the spin-incoherent case.  Of course, from Eq.(\ref{eq:I_mobile_final}) it is clear that the effects of $\delta_a$ are still felt, but only as a prefactor to the frequency dependence.

The effects of an externally applied magnetic field can also be easily incorporated \cite{Fiete:prl06}.  For a quantum wire the magnetic field does not affect the orbital part of the wavefunction until the magnetic length is of the order of the wire width.  We therefore only discuss the coupling of the magnetic field to the spin of the electron, {\it i.e.} a Zeeman coupling.  The magnetic field will influence the spin correlations described by $c^+_\sigma(q)$.  In the spin-incoherent regime with zero field one has $\langle Z_{l,\sigma} Z^\dagger_{0,\sigma}\rangle = 2^{-|l|}$, independent of $\sigma$.  In a finite external field one instead has $\langle Z_{l,\sigma} Z^\dagger_{0,\sigma}\rangle = p_\sigma^{|l|}$ where $p_\uparrow=(1+\exp\{-E_Z/k_BT\})^{-1}=1-p_\downarrow$ and $E_Z$ is the Zeeman energy \cite{Fiete:prl06,Kindermann_crossover:prb06}.  While the presence of the magnetic field may shift the threshold energy, it will not affect the exponent Eq.(\ref{eq:a_SILL_mobile}) \cite{Fiete:prl06}.  The magnetic field also gives rise to a frequency scale $\omega^*=\frac{E_c}{\hbar}e^{-2E_Z/(gk_B T)}$ below which SILL behavior is observed and above which behavior characteristic of a spinless LL is observed \cite{Fiete:prl06,Kindermann_crossover:prb06}, namely the form Eq.(\ref{eq:I_FL}) with the exponent Eq.(\ref{eq:a_LL_mobile}).

We next turn to the case of an infinitely massive impurity for which backscattering is irrelevant.  In this case the effective electron-hole interaction is given by Eq.(\ref{eq:e-h}) only with $\delta_a=0$ as this corresponds to the case of a stationary hole with no backscattering.  The photon absorption in this case is thus given by Eq.(\ref{eq:I_mobile_final}) with  $\delta_a=0$.  While an infinite mass hole breaks translational symmetry, the irrelevance of backscattering effectively restores translational symmetry at low energies and the threshold behavior maps onto the finite mass case.

On the other hand, when backscattering is relevant the situation is very different.  At low energies the hole acts as a boundary to a semi-infinite one dimensional system \cite{Kane:prl92,Furusaki:prb93}. In this case, the SILL maps onto a {\em spinless} electron system \cite{Fiete_2:prb05,Fiete:rmp07} which allows us to immediately exploit \cite{Fiete:prl06} the results obtained for that system.  In particular, one obtains a universal contribution of 1/8 to the threshold exponent \cite{Kane:prb94,Prokofev:prb94,Affleck:jpa94,Furusaki:prb97,Komnik:prb97},
\begin{equation}
\alpha^{SILL}_{\rm immobile, relevant}=1-\frac{1}{g}(1-\delta_s)^2-1/8,
\end{equation}
and the frequency dependence follows Eq.(\ref{eq:I_FL}) {\em without} the logarithmic corrections. In fact, for the infinitely massive hole with relevant backscattering the only signature of spin-incoherent physics is hidden in the mapping $g=2g_c$ \cite{Fiete_2:prb05}.

\section{Outstanding issues in the spin-incoherent Luttinger liquid}

While our theoretical understanding of the spin-incoherent regime has advanced rapidly, the difficulty of obtaining the regime $E_s \ll E_c$ has hampered experimental tests of the theory.  To date our best experimental evidence comes from momentum resolved tunneling on cleaved-edge overgrowth quantum wires \cite{Steinberg:prb06}, transport in split-gate devices \cite{Hew:prl08}, and somewhat more speculatively in transport measurements on gated single wall carbon nanotubes with low electron density \cite{Deshpande:np08}.  Given that the existing theory now encompasses hybrid structures of SILLs with ferromagnets and superconductors as well \cite{Tilahun:prb08,Tilahun:prb09}, the number of experimental groups poised to make contributions to this exciting field is greatly enlarged.  For a recent discussion of numerous possible experiments that would provide strong evidence for spin-incoherent Luttinger liquid behavior see \cite{Fiete:rmp07}.

On the theoretical side, a number of open issues remain. In particular, it is desirable to have a better understanding of  the behavior on temperature scales $k_B T \approx E_s$ which is most likely very relevant for many experiments that approach but do not quite reach the spin-incoherent regime.  Related to this is the need for  a better understanding of the crossover between the Luttinger liquid and the spin-incoherent Luttinger liquid regimes.  Both of these issues will likely require a numerical attack as there are no obvious analytical methods available to address them.  Finally there is the issue of spin-orbit coupling that has so far received no attention.  For very strong spin-orbit coupling is there novel behavior in the regime $E_s \ll k_B T \ll E_{SO},E_c$?  These issues await further study.

\ack
I am grateful to my collaborators Ophir Auslaender, Leon Balents, Adrian Feiguin, Thierry Giamarchi, Bert Halperin, Anibal Iucci,  Markus Kindermann, Karyn Le Hur, Jiang Qian, Hadar Steinberg, Dagim Tilahun, Yaroslav Tserkovnyak, and Amir Yacoby for their important contributions to our work on spin-incoherent Luttinger liquids.  I am also indebted to Matthew Fisher, Akira Furusaki, Leonid Glazman, and Kostya Matveev for many enlightening discussions on this topic.  This work was supported by NSF Grant No. PHY05-51164 and the Lee A. DuBridge Foundation.

\section*{References}



\end{document}